# Generative MR Multitasking with complex-harmonic cardiac encoding: Bridging the gap between gated imaging and real-time imaging

Xinguo Fang[1,2], Anthony G. Christodoulou[1,2*]

[1] Department of Radiological Sciences, David Geffen School of Medicine at UCLA, Los Angeles, California, USA

[2] Department of Bioengineering, University of California, Los Angeles, Los Angeles, California, USA

* Correspondence to:

Anthony G. Christodoulou, PhD

300 UCLA Medical Plaza Suite B119

Los Angeles, CA 90095

Email: AChristodoulou@mednet.ucla.edu

**Abstract**

**Purpose:** To develop a unified image reconstruction framework that bridges real-time and gated cardiac MRI, including quantitative MRI.

**Methods:** We introduce Generative Multitasking, which learns subject- and dataset-specific implicit neural temporal bases from sequence timings and an interpretable latent space for cardiac and respiratory motion. Cardiac motion is modeled as a complex harmonic, with phase encoding timing and a latent amplitude capturing beat-to-beat functional variability, linking cardiac phase-resolved ("gated-like") and time-resolved ("real-time-like") views. We implemented the framework using a conditional variational autoencoder (CVAE) and evaluated it for free-breathing, non-ECG-gated radial GRE in three settings: steady-state cine imaging, multicontrast T2prep/inversion-recovery imaging, and dual-flip-angle T1/T2 mapping, compared with conventional Multitasking.

**Results:** Generative Multitasking provided flexible cardiac motion representation, enabling reconstruction of archetypal cardiac phase-resolved cines (like gating) as well as time-resolved series that reveal beat-to-beat variability (like real-time imaging). Conditioning on the previous k-space angle and modifying this term at inference removed eddy-current artifacts without globally smoothing high temporal frequencies. For quantitative mapping, Generative Multitasking reduced intraseptal T1 and T2 coefficients of variation (CoV) compared with conventional Multitasking (T1: 0.13 vs. 0.31; T2: 0.12 vs. 0.32; $p < 0.001$), indicating higher SNR.

**Conclusion:** Generative Multitasking uses a CVAE with complex harmonic cardiac coordinates to unify gated and real-time CMR within a single free-breathing, non-ECG-gated acquisition. The framework allows flexible cardiac motion representation, suppresses trajectory-dependent artifacts, and improves T1 and T2 mapping, suggesting a path toward cine, multicontrast, and quantitative imaging without separate gated and real-time scans.

Abstract word count (exclude Keywords): 240

## 1. Introduction:

Cardiovascular magnetic resonance imaging (CMR) is a powerful source of information that can aid in patient management. However, compared to MRI of other body parts, CMR faces specific challenges from cardiac and respiratory motion. Broadly, there are two approaches to handling cardiac motion: cardiac-phase-resolved imaging (i.e., gated or segmented imaging) and time-resolved (i.e., "real-time" imaging)[1].

In gated sequences, data are reorganized or triggered according to the estimated phase of the cardiac cycle, either determined from external sensors such as an ECG or pilot tone, or directly from imaging data ("self-gating"). However, cardiac phase binning captures only differences in timing without accounting for functional variability across heartbeats. As a result, irregular motion or functional variability lead to reconstruction artifacts caused by inconsistent data across multiple heartbeats. Patients with arrhythmias remain inadequately served by gated CMR techniques, with image quality severely compromised due to multiple ectopic beats or atrial fibrillation[2]. Cardiac phase representation is inherent both to cine imaging and a central part of quantitative imaging, as quantification of properties such as T1 and T2 requires aggregating images from multiple heart beats. Thus, promising new imaging frameworks such as "all-in-one" frameworks[3] using multidimensional imaging to capture and analyzing tissue properties and motion are generally cardiac phase-gated.

Real-time cardiovascular magnetic resonance (CMR) aims for time-resolved imaging with high frame rates, to separately depict multiple individual heartbeats. Most commonly, real-time CMR employs single-shot imaging, avoiding directly sharing data across multiple heart beats. In patients with irregular rhythms, real-time imaging reduces the need for repeat scanning due to failed gating[4]. However, achieving high temporal resolution while forgoing cardiac phase segmentation generally comes at the cost of spatial resolution, so real-time CMR is generally limited in spatial resolution, and quantification of tissue properties measured across multiple heart beats with real-time CMR is complicated by the lack of a cardiac phase definition.

Conventionally, the gated imaging CMR framework and "real-time" CMR framework are considered mutually exclusive. Here we propose a missing link between cardiac phase-resolved frameworks (e.g., gated or all-in-one CMR) and time-resolved frameworks (e.g., real-time CMR): a latent amplitude representation of the cardiac cycle that encodes beat-to-beat variation. While cardiac phase continues to encode timing similarities and differences across cycles, the amplitude dimension captures functional variability from one beat to the next. Together, this complex harmonic representation provides the structure of defined cardiac phases, like gated/segmented imaging, without requiring that each cardiac cycle is the same, like real-time imaging.

In this work, we develop and evaluate a generative Multitasking framework that embeds this complex harmonic cardiac representation in a subject- and dataset-specific implicit neural temporal model. A conditional variational autoencoder (CVAE) with a structured, interpretable latent space performs both motion identification (self-gating) and temporal subspace generation from arbitrary locations in that latent space. Within this framework, the same acquisition can be reconstructed either as a cardiac phase-resolved, "gated-like" multidimensional dataset for an archetypal cardiac beat or as a time-resolved, "real-time-like" series that retains beat-to-beat variability. We apply this framework to three representative use cases under free-breathing, non-ECG-gated conditions: steady-state cine imaging, multicontrast T2prep/inversion-recovery imaging, and dual-flip-angle T1/T2 mapping, to assess the framework's ability to represent motion, handle trajectory-dependent system effects, and support quantitative mapping, respectively.

## 2. Methods:

### 2.1. Conventional Multitasking Image Model

Multitasking[5] reconstructs a high-dimensional image $A(\mathbf{x}, \boldsymbol{\tau})$ a function of spatial location $\mathbf{x}$ and of $N$ time-varying independent variables ("time dimensions") collected as $\boldsymbol{\tau} = [\tau_1 \; \tau_2 \; \cdots \; \tau_N]^T = [\boldsymbol{\tau}_{\mathcal{M}}^T \; \boldsymbol{\tau}_{\mathcal{C}}^T]^T$. Here $\boldsymbol{\tau}_{\mathcal{M}}$ comprises motion-related time dimensions whose chronological timings $\boldsymbol{\tau}_{\mathcal{M}}[t]$ are unknown *a priori* (e.g., cardiac and respiratory states,), and $\boldsymbol{\tau}_{\mathcal{C}}$ collects dimensions that describe contrast-related time dimensions determined by pulse-sequence timings and parameters (e.g., inversion time, T2prep duration, flip angle), whose chronological timings $\boldsymbol{\tau}_{\mathcal{C}}[t]$ are known by design. Multitasking uses a subspace model to represent $A(\mathbf{x}, \boldsymbol{\tau})$ as the product of an $L$-dimensional spatial factor $\mathbf{u}(\mathbf{x})$ and an $L$-dimensional temporal factor $\boldsymbol{\varphi}(\boldsymbol{\tau})$:

$$A(\mathbf{x}, \boldsymbol{\tau}) = \mathbf{u}^T(\mathbf{x})\boldsymbol{\varphi}(\boldsymbol{\tau}) = \sum_{\ell=1}^{L} u_\ell(\mathbf{x})\varphi_\ell(\boldsymbol{\tau}) \quad (1)$$

The standard Multitasking approach specifically models $\boldsymbol{\varphi}(\boldsymbol{\tau})$ as a discrete, separable low-rank tensor:

$$\varphi_\ell(\tau_1, \tau_2, \cdots, \tau_N) = \sum_{\ell_1}^{L_1}\sum_{\ell_2}^{L_2}\cdots\sum_{\ell_N}^{L_N} c_{\ell\ell_1\ell_2\cdots\ell_N} v_{1,\ell_1}[\tau_1] v_{2,\ell_2}[\tau_2]\cdots v_{N,\ell_N}[\tau_N] \quad (2)$$

where $c_{\ell\ell_1\ell_2\cdots\ell_N}$ indexes elements of a core tensor, and where $\{v_{n,\ell_n}[\tau_n]\}_{\ell_n=1}^{L_n}$ is a discrete-time basis for the *n*th temporal dimension. Multitasking image reconstruction (Fig. 1A) comprises three main steps:

1. Identify the motion timings $\boldsymbol{\tau}_{\mathcal{M}}[t]$ from time-resolved self-gating data, in order to gate/bin the cardiac and respiratory data into the multidimensional space. This is typically performed by Bloch-equation aware clustering.
2. Estimate the temporal factor $\boldsymbol{\varphi}(\boldsymbol{\tau})$ from those same self-gating data, typically by low-rank tensor completion of the binned self-gating data, followed by a high-order SVD to calculate a tensor subspace.
3. Estimate the spatial factor $\mathbf{u}(\mathbf{x})$, typically by regularized least-squares fitting of the estimated $\boldsymbol{\varphi}(\boldsymbol{\tau})$ to the full raw data.

### 2.2. Generative Multitasking Image Model

Generative Multitasking (Fig. 1B) replaces the low-rank tensor model with an implicit neural representation[6-8] of $\boldsymbol{\varphi}(\boldsymbol{\tau})$:

$$\boldsymbol{\varphi}(\boldsymbol{\tau}) = D(\boldsymbol{\tau}_{\mathcal{M}}, \boldsymbol{\tau}_{\mathcal{C}}) \quad (3)$$

where $D(\cdot)$ is, e.g. the decoder of a subject- and dataset-specific conditional variational autoencoder (CVAE). The decoder is paired with an encoder that converts time-resolved input data $\boldsymbol{\varphi}[t]$ and conditional inputs $\boldsymbol{\tau}_{\mathcal{C}}[t]$ into motion timings $\boldsymbol{\tau}_{\mathcal{M}}[t]$, thus also performing motion identification. The CVAE therefore replaces both the motion identification (binning) and temporal factor estimation (tensor subspace estimation) steps of Multitasking reconstruction by the encoder and decoder, respectively. The decoder can be queried either at the original acquired time points, $\boldsymbol{\varphi}(\boldsymbol{\tau}[t]) = D(\boldsymbol{\tau}_{\mathcal{M}}[t], \boldsymbol{\tau}_{\mathcal{C}}[t])$, for time-resolved "real-time-like" image generation or at arbitrary $(\boldsymbol{\tau}_{\mathcal{M}}, \boldsymbol{\tau}_{\mathcal{C}})$ combinations for multidimensional "gated-like" image generation.

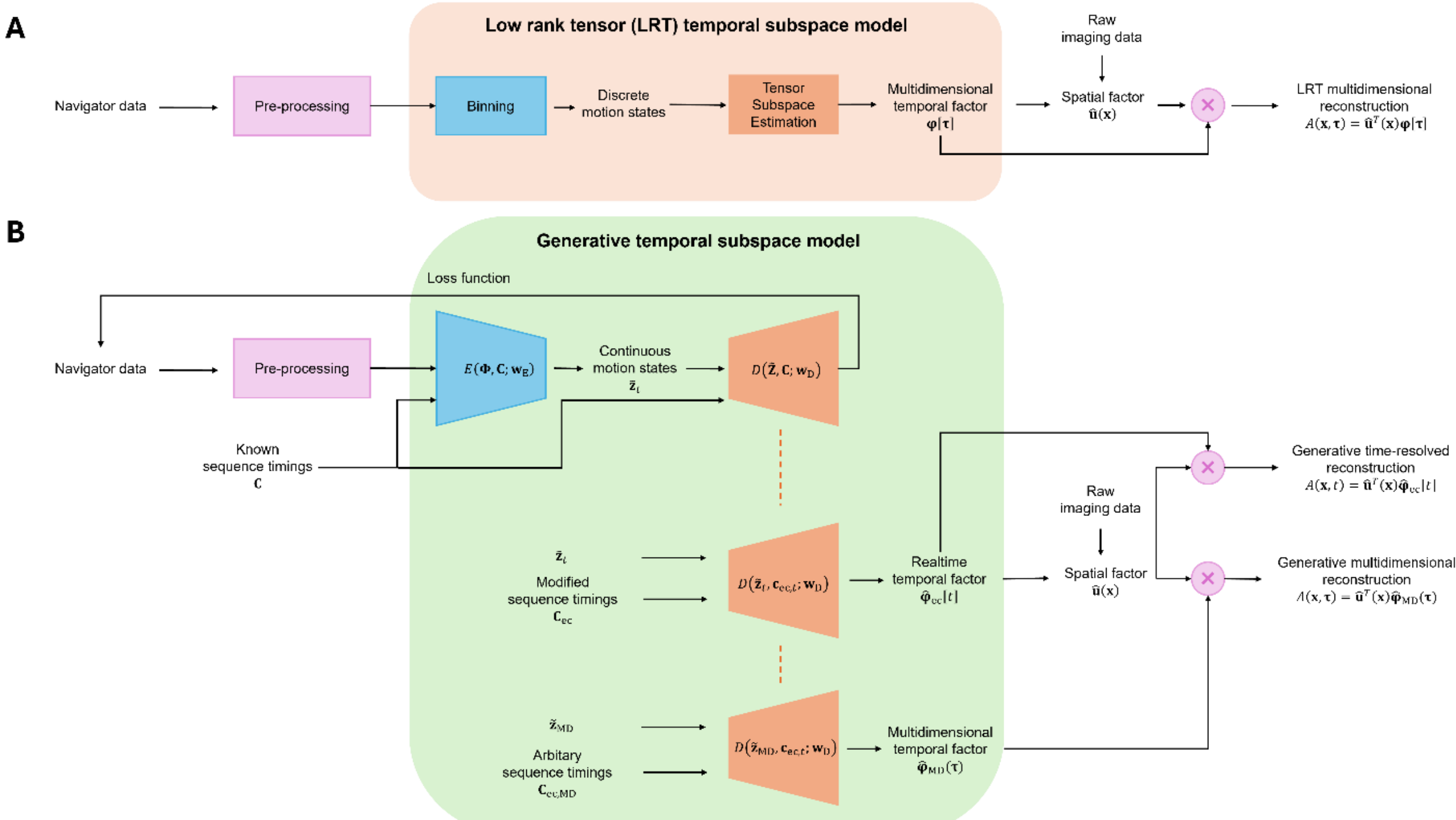

**Figure 1**: Comparison of the conventional (A) and generative (B) Multitasking temporal modeling processes. Generative Multitasking learns a scan- and subject-specific latent space which is directly interpretable as motion dimensions, such that an encoder can replace cardiac and respiratory binning, and a decoder can replace tensor subspace estimation with an implicit neural representation. Once the network is trained, the decoder can be used to generate: 1) an eddy-current corrected time-resolved temporal factor to help calculate the spatial factor, and 2) temporal factors for arbitrary motion/pulse sequence timing combinations which allow switching between various gated-like and real-time-like image display modes.

## 2.3. Conditional Variational Autoencoder (CVAE)

The core module in our study is a CVAE[9] (Fig. 2), a deep conditional generative model. Here, our CVAE ingests time-resolved inputs $\boldsymbol{\varphi}[t]$, samples of a distribution $\varphi$, alongside conditional inputs $\boldsymbol{\tau}_{\mathcal{C}}[t]$, samples of a distribution $c$, to reconstruct time-resolved outputs $\hat{\boldsymbol{\varphi}}[t]$. Conditioning on the pulse sequence timings $\boldsymbol{\tau}_{\mathcal{C}}[t]$ alerts the network to contrast weighting changes due to e.g., preparation pulses, so that the learned latent representation $z$ captures motion rather than contrast variations. Between the encoder $E(z|\varphi, c)$ and decoder $D(\hat{\varphi}|z, c)$, we insert a filter bank and harmonic analysis module $f(\cdot)$ to encourage $z$ to be directly interpretable as cardiac and respiratory motion states.

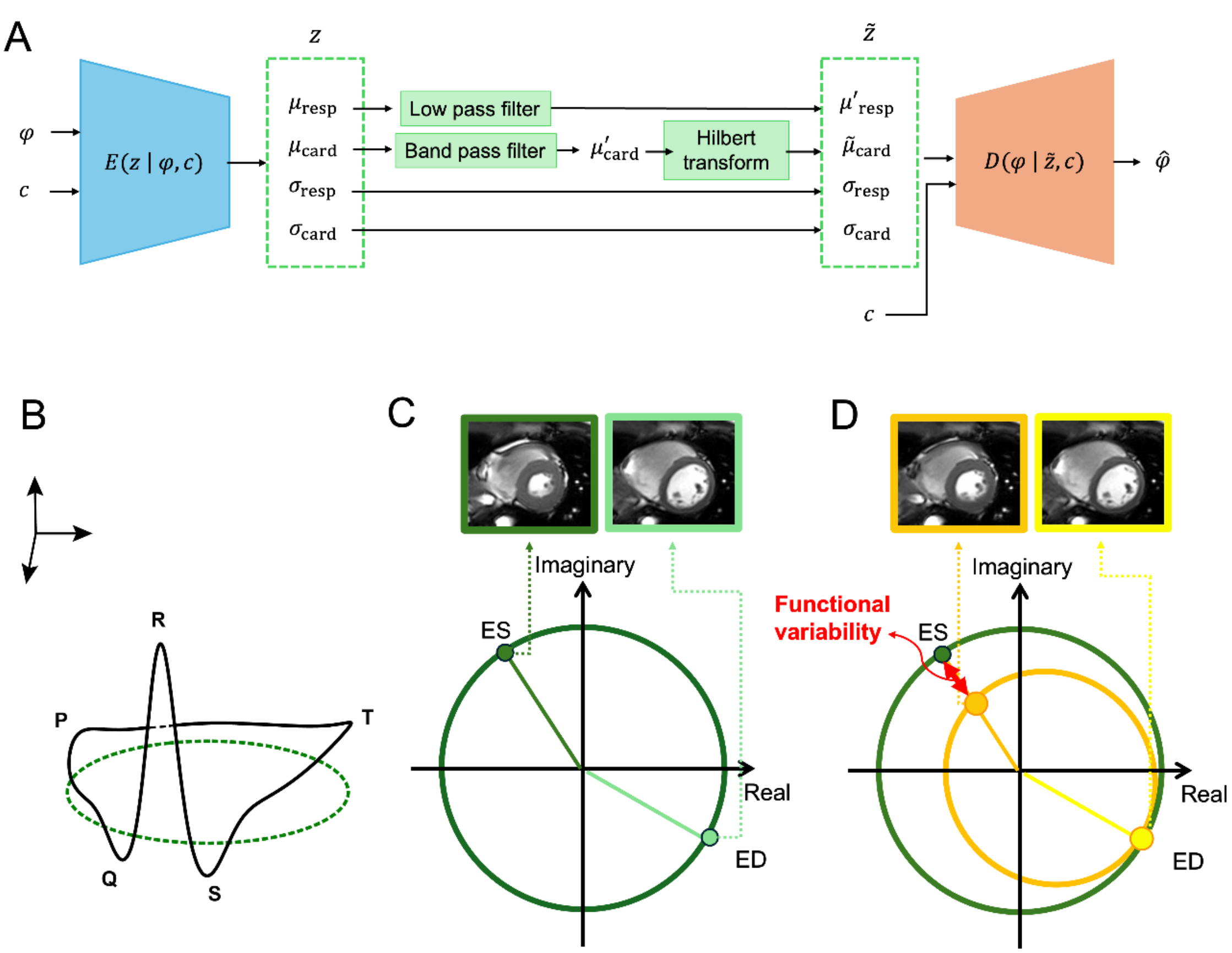

**Figure 2:** Diagram of the conditional variational autoencoder (CVAE) and conceptual illustration of the complex harmonic cardiac representation inside the latent space. (A) Time-resolved samples of an input distribution $\varphi$ and conditional inputs $c$ containing pulse sequence timings are encoded into a latent space $z$. The latent space is enforced to be interpretable as motion states: a filter bank disentangles respiratory and cardiac latent vectors, and a Hilbert transform turns a real-valued cardiac latent vector into a complex-harmonic cardiac latent vector with variable amplitude which preserves beat-to-beat differences in motion states. The decoder takes the processed latent vectors and the conditional $c$ to reproduce the input $\varphi$. (B) Conceptual illustration of an ECG signal, typically used to identify cardiac phase, as lying along a unit-amplitude complex harmonic. (C) Illustration of the locations of cardiac images at different phases along unit-amplitude complex harmonic, with phase capturing only the timing of the cardiac cycle. (D) Illustration of adding a variable latent "amplitude" to the complex harmonic in order to encode cardiac cycles with beat-to-beat functional variability.

### 2.3.1. Encoder inputs and outputs

Our study implemented the learned encoder as a nonlinear time-invariant mapping from time-resolved inputs to latent motion coordinates, $\mathbf{z}_t = E(\boldsymbol{\varphi}_t, \mathbf{c}_t; \mathbf{w}_\mathrm{E})$. Each vector $\boldsymbol{\varphi}_t = \boldsymbol{\varphi}[t]$ contains self-gating data, an initial time-resolved reconstruction, features of either type of input, or other time-resolved motion surrogate. The conditional vector $\mathbf{c}_t = [\boldsymbol{\tau}_C^T[t] \quad k[t - T_R]]^T$ includes the known pulse sequence timings $\boldsymbol{\tau}_C[t]$ as well as the k-space coordinates of the immediately preceding readout, $k[t - T_R]$, so the network can model eddy current artifacts from k-space jumps prior to each self-gating readout[10-13]. The $\mathbf{w}_\mathrm{E}$ are learned subject -and scan-specific network weights. The encoder outputs the time-varying distribution parameters (mean and log-variance) of a two-dimensional latent space encoding respiratory and cardiac motion:

$$\mathbf{z}_t = \begin{bmatrix} \mathbf{z}_{\mathrm{resp},t} \\ \mathbf{z}_{\mathrm{card},t} \end{bmatrix} = \begin{bmatrix} \mu_\mathrm{resp}[t] \\ \log \sigma_\mathrm{resp}^2 [t] \\ \mu_\mathrm{card}[t] \\ \log \sigma_\mathrm{card}^2 [t] \end{bmatrix} \quad (4)$$

### 2.3.2. Cardiac and respiratory disentanglement

During training, we enforce separation of respiratory and cardiac dynamics by assembling the latent means at all time points into time-resolved coordinates $\mu_\mathrm{resp}[t]$ and $\mu_\mathrm{card}[t]$ and applying frequency-based disentanglement:

- $\mu'_\mathrm{resp}[t]$: output of a low-pass filter on $\mu_\mathrm{resp}[t]$
- $\mu'_\mathrm{card}[t]$: output of a band-pass filter to $\mu_\mathrm{card}[t]$

This frequency-domain disentanglement leverages physiological distinctions between respiratory motion (typically slower and lower frequency) and cardiac motion (higher frequency), thus refining the specificity of each latent vector.

### 2.3.3. Complex harmonic cardiac representation

Between the band-pass filter and decoder, a Hilbert transform generates a complementary imaginary component to the real-valued cardiac latent signal $\mu'_\mathrm{card}[t]$ to form the complex-valued, analytic signal $\tilde{\mu}_\mathrm{card}[t]$, a complex-harmonic representation of cardiac motion. The complex representation encodes cardiac phase ($\angle\tilde{\mu}_\mathrm{card}[t]$) as in standard cardiac phase representations (e.g., gating), but also contains a latent time-varying amplitude ($|\mu_{\text{card}}[t]|$) in order to capture beat-to-beat characteristics as in real-time imaging. (Fig. 3). Figure S1 illustrates how the cardiac bandpass filter and Hilbert transform (the only time-dependent modules) affect signals with varying timing and amplitude irregularities.

### 2.3.4. Decoder inputs and outputs

We constructed the learned decoder in the CVAE as another nonlinear time-invariant system which now converts latent space motion coordinates back into the original time-resolved inputs. We implemented this as a neural network $\hat{\boldsymbol{\varphi}}_t = D(\tilde{\mathbf{z}}_t, \mathbf{c}_t; \mathbf{w}_{\text{D}})$ with learned scan-specific network weights $\mathbf{w}_{\text{D}}$. At each timepoint *t* it receives three latent inputs: the first sampled from the real-valued time-varying distribution $z_{\text{resp}}[t] \sim \mathcal{N}\left(\mu'_{\text{resp}}[t], \sigma^2_{\text{resp}}[t]\right)$ and the other two sampled from the real and imaginary components of the complex-valued time-varying distribution $\tilde{z}_{\text{card}}[t] \sim \mathcal{N}\left(\tilde{\mu}_{\text{card}}[t], \sigma^2_{\text{card}}[t]\right)$:

$$\tilde{\mathbf{z}}_t = \begin{bmatrix} z_{\text{resp}}[t] \\ Re(\tilde{z}_{\text{card}}[t]) \\ Im(\tilde{z}_{\text{card}}[t]) \end{bmatrix} \quad (5)$$

alongside the conditional inputs $\mathbf{c}_t$ to reconstruct the original inputs as $\hat{\boldsymbol{\varphi}}_t$.

As a whole, the CVAE performs:

$$\hat{\boldsymbol{\Phi}} = CVAE(\boldsymbol{\Phi}, \mathbf{C}; \mathbf{w}_{\text{E}}, \mathbf{w}_{\text{D}}) = D(f[E(\boldsymbol{\Phi}, \mathbf{C}; \mathbf{w}_{\text{E}})], \mathbf{C}; \mathbf{w}_{\text{D}}). \quad (6)$$

Here, $\boldsymbol{\Phi}$ and $\mathbf{C}$ are the concatenations of the full time series; their *t*th columns are $\boldsymbol{\varphi}_t$ and $\mathbf{c}_t$, respectively. The encoder $E(\boldsymbol{\Phi}, \mathbf{C}; \boldsymbol{w}_{\text{E}})$ and decoder $D(\tilde{\mathbf{Z}}, \mathbf{C}; \boldsymbol{w}_{\text{D}})$ accept the full time series. The function $f(\cdot)$ comprises the cardiac/respiratory disentanglement filter bank and Hilbert transformation that give the latent space its interpretable structure. In our implementation, the encoder and decoder operate independently on each time point; only $f(\cdot)$ operates along the time domain.

### 2.3.5. Training loss function

For each dataset, the network is trained by minimizing a cost function which balances reconstruction fidelity with latent space regularization:

$$\arg\min_{\{\boldsymbol{w}_{\mathrm{E}},\boldsymbol{w}_{\mathrm{D}}\}} \|\boldsymbol{\Phi} - CVAE(\boldsymbol{\Phi},\mathbf{C};\boldsymbol{w}_{\mathrm{E}},\boldsymbol{w}_{\mathrm{D}})\|^2 + \beta \cdot KL[E(\boldsymbol{\Phi},\mathbf{C};\boldsymbol{w}_{\mathrm{E}})\| \mathcal{N}(\mathbf{0},\mathbf{I})] \quad (7)$$

The first term measures reconstruction fidelity using the expected log-likelihood of the data assuming white Gaussian noise in $\boldsymbol{\Phi}$. The second term is the Kullback–Leibler (KL) divergence between the samples of the posterior distribution $E(z|\varphi, c)$ and the prior distribution $p(z) = \mathcal{N}(0, I)$; the hyperparameter $\beta$ controls the degree of regularization. During training, the inputs to the decoder (after filtering and Hilbert transformation) are reparameterized to further sample the distribution of the input latent space and promote regularity.

#### 2.3.6. Inference

After the network is trained, the decoder can generate temporal representations at arbitrary combinations of latent space coordinates $\tilde{\mathbf{z}}$ and conditional inputs $\mathbf{c}$. Because the latent space locations and conditional inputs are interpretable as motion states and pulse sequence timings, respectively, this permits generation of $\widehat{\boldsymbol{\varphi}} = D(\tilde{\mathbf{z}}, \mathbf{c}; \mathbf{w}_{\mathrm{D}})$ at arbitrary motion state/pulse sequence timing combinations, a key feature of high-dimensional imaging with multiple time dimensions.

### 2.4. Generative Multitasking Image Calculation

#### 2.4.1. Image calculation

Prior to CVAE learning, we calculate an initial time-resolved image by time-resolved subspace-constrained imaging. We use the right singular vectors $\mathbf{V}$ to define the subspace, calculating an initial spatial factor

$$\widehat{\mathbf{U}}_0 = \arg\min_{\mathbf{U}} \|\mathbf{d} - \Omega(\mathbf{FSUV}^H)\|_2^2. \quad (8)$$

Here, $\mathbf{F}$ is the Fourier transform, $\mathbf{S}$ applies sensitivity maps, and $\Omega(\cdot)$ performs k-t space undersampling. Calculating $\boldsymbol{\Phi} = \left(\widehat{\mathbf{U}}_0^H \widehat{\mathbf{U}}_0\right)^{1/2} \mathbf{V}^H$ defines "image-weighted" time-resolved features $\boldsymbol{\varphi}_t$ from the $t$th column of $\boldsymbol{\Phi}$.

We train the CVAE on $\boldsymbol{\varphi}_t$ and use the trained decoder to generate a time-resolved, eddy-corrected temporal factor $\boldsymbol{\Phi}_{\text{ec}}$ with columns $\boldsymbol{\varphi}_{\text{ec},t} = D\left(\begin{bmatrix} {\mu'}_{\text{resp}}[t] \\ Re(\tilde{\mu}_{\text{card}}[t]) \\ Im(\tilde{\mu}_{\text{card}}[t]) \end{bmatrix}, \mathbf{c}_{\text{ec},t}; \mathbf{w}_{\text{D}}\right)$ constructed from the learned, deterministic timings $\boldsymbol{\tau}_{\mathcal{M}}[t] = [\tau_{\text{resp}}[t] \quad \tau_{\text{card}}[t]]^T = \left[{\mu'}_{\text{resp}}[t] \quad \tilde{\mu}_{\text{card}}[t]\right]^T$ (i.e., $\tilde{\mathbf{z}}_t$ with latent space variances $\sigma^2$ set to zero), and a modified conditional $\mathbf{c}_{\text{ec},t} = [\boldsymbol{\tau}_{\mathcal{C}}^T[t] \quad k_{\text{SG}}]^T$ in which entries that originally encoded the immediately preceding k-space trajectory, $k[t - T_R]$, are replaced by the constant coordinates of the self-gating lines, $k[t] = k_{\text{SG}}$. This allows the decoder to regenerate the data as if no k-space jump occurred, effectively removing eddy-current artifacts.

Given $\boldsymbol{\Phi}_{\text{ec}}$, we calculate a discretized spatial factor $\mathbf{U}$, whose *i*th row equals $\mathbf{u}^T(\mathbf{x}_i)$, as a least-squares fit to the acquired k-t space data $\mathbf{d}$:

$$\widehat{\mathbf{U}} = \arg\min_{\mathbf{U}} \|\mathbf{d} - \Omega(\mathbf{FSU}\boldsymbol{\Phi}_{\text{ec}})\|_2^2 \qquad (9)$$

Note that it is straightforward to add a regularizer to the least-squares objective, but in this work we limit our results to the unregularized solution.

Once $\widehat{\mathbf{U}}$ is known, one can generate various eddy-corrected time-resolved images using the same spatial factor $\widehat{\mathbf{u}}^T(\mathbf{x})$ for all reconstruction variations, varying only the decoder inputs in order to traverse the multidimensional parameter space, such as

- Time-resolved images: $A_{\text{tr}}(\mathbf{x}, t) = \widehat{\mathbf{u}}^T(\mathbf{x}_i) D\left(\begin{bmatrix} {\mu'}_{\text{resp}}[t] \\ Re(\tilde{\mu}_{\text{card}}[t]) \\ Im(\tilde{\mu}_{\text{card}}[t]) \end{bmatrix}, \mathbf{c}_{\text{ec},t}; \mathbf{w}_{\text{D}}\right)$
- Time-resolved images frozen at respiratory state $\tau_{\text{resp},0}$: $A_{\text{tr}}\left(\mathbf{x}, t; \tau_{\text{resp},0}\right) = \widehat{\mathbf{u}}^T(\mathbf{x}_i) D\left(\begin{bmatrix} \tau_{\text{resp},0} \\ Re(\tilde{\mu}_{\text{card}}[t]) \\ Im(\tilde{\mu}_{\text{card}}[t]) \end{bmatrix}, \mathbf{c}_{\text{ec},t}; \mathbf{w}_{\text{D}}\right)$

- Arbitrary multidimensional images $A(\mathbf{x}, \boldsymbol{\tau}) = \hat{\mathbf{u}}^T(\mathbf{x}_i) D\left(\begin{bmatrix}\tau_{\text{resp}} \\ Re(\tau_{\text{card}}) \\ Im(\tau_{\text{card}})\end{bmatrix}, \begin{bmatrix}\boldsymbol{\tau}_{\mathcal{C}} \\ k_{\text{SG}}\end{bmatrix}; \mathbf{w}_{\text{D}}\right)$
- Cardiac-phase resolved multidimensional images for an "archetypal" cardiac cycle $\tau_{\text{card}}(\theta) = \bar{a}(\theta)e^{i\theta}$: $A_{\text{cycle}}\left(\mathbf{x}, \tau_{\text{resp}}, \theta_{\text{card}}, \boldsymbol{\tau}_{\mathcal{C}}\right) = \hat{\mathbf{u}}^T(\mathbf{x}_i) D\left(\begin{bmatrix}\tau_{\text{resp}} \\ \bar{a}(\theta)\cos(\theta) \\ \bar{a}(\theta)\sin(\theta)\end{bmatrix}, \begin{bmatrix}\boldsymbol{\tau}_{\mathcal{C}} \\ k_{\text{SG}}\end{bmatrix}; \mathbf{w}_{\text{D}}\right)$

In the following experiments we use this generative decoder both to reconstruct acquired real-time data and to synthesize multidimensional and archetypal cardiac phase-resolved image series.

### 2.5. Experiments

We evaluated the framework using three pulse sequences of increasing complexity. First, we demonstrated the cardiac motion representation in steady-state cine imaging. Second, we added T2prep/inversion-recovery (T2IR) preparation pulses to demonstrate eddy current removal and contrast–motion disentanglement. Third, we applied a multi-flip-angle T1/T2 mapping sequence to demonstrate and measure the impact on quantitative mapping. Images were reconstructed using conventional Multitasking, which uses discrete cardiac phase binning (i.e., gating) and also with generative Multitasking, with its continuous complex harmonic motion representation.

All scans were performed at 3T using continuous-acquisition radial GRE with interleaved self-gating lines at the 0° radial spoke. For all scans, the field of view was 270 mm × 270 mm, spatial resolution was 1.7 mm × 1.7 mm × 8 mm, and TR/TE were 3.6/1.6 ms. No fat suppression was used. Scans were performed in 2D short-axis orientation. The spoke angle $\angle k$ fully parameterizes the radial k-t space sampling schedule, so we defined the k-space trajectory conditional input as $\angle k[t - T_R]$ rather than paired $\left(k_x, k_y\right)$-coordinates. All scans were free-breathing and without ECG gating.

*Cine demonstration*. We first demonstrated the complex cardiac motion representation using a steady-state GRE cine acquisition with flip angle (FA)=12° in a healthy volunteer. We used a tiny-golden-angle sampling scheme with self-gating lines every 16th readout (every 57 ms).

Because there were no preparation pulses, the only conditional input was $\mathbf{c}_t = \angle k[t - T_R]$. Acquisition time was 1 min.

<u>*Multicontrast demonstration*</u>. Next, we reconstructed T2IR-prepared GRE data from *N*=3 subjects to demonstrate eddy-current artifact removal and contrast–motion disentanglement. We evaluated eddy-current artifacts because this analysis isolates how the CVAE affects the temporal basis before those effects propagate to the final quantitative maps, and because RF and gradient history effects are well-known confounders in self-gating data[10-13]. T2IR preparation times cycled between $\tau_{\text{prep}}$ = 0 (a pure IR pulse), 30, 40, 50, and 60 ms; readouts used FA=5°. We used a golden-angle sampling scheme with self-gating lines every other readout (every 7.1 ms). The conditional vector was $\mathbf{c}_t = \begin{bmatrix} T_I[t] \\ \tau_{\text{prep}}[t] \\ \angle k[t - T_R] \end{bmatrix}$, where $T_I$ is the inversion time. Acquisition time was 1.5 min.

<u>*Quantitative mapping evaluation*</u>. We subsequently applied the framework to a dual-flip-angle, simultaneous multi-slice (SMS) Multitasking acquisition in *N* = 7 healthy volunteers. The sequence imaged three short-axis slices (base, mid, and apex). The T2IR modules and sampling scheme were as above; FA alternated between 3° and 10° to enable joint B1+/T1/T2 mapping. The conditional vector was $\mathbf{c}_t = \begin{bmatrix} T_I[t] \\ \tau_{\text{prep}}[t] \\ FA[t] \\ \angle k[t - T_R] \end{bmatrix}$. Acquisition time was 3 min.

We calculated T1 and T2 maps from images at the median-amplitude end-diastole cardiac coordinate and end-expiratory respiratory coordinate. For each slice, we computed the mean and standard deviation of voxelwise T1 and T2 values within the myocardial septum and used these to calculate the intraseptal coefficient of variation (CoV)=$\sigma/\mu$, i.e., the inverse of SNR. We performed a two-way ANOVA to control for multiple slices when comparing CoV between methods, with a significance level of 0.05. For both Multitasking and generative Multitasking, we did not add spatial regularization to Eq. 9, in order to isolate the impact of the low-rank tensor vs. implicit neural representation–generated temporal subspaces and to preserve linear reconstruction for more independent noise measurements.

To measure the impact of our network on reconstruction time, we recorded network training and inference times and compared them with the times required for binning and tensor subspace estimation in conventional Multitasking (the two steps which the network replaces). All computations were performed on a Linux workstation with 1 TB RAM, two 36-core Intel Xeon Platinum CPUs, and two NVIDIA RTX 6000 Ada Generation GPUs. We compared computation times using a Wilcoxon signed-rank test, regarding $p<0.05$ as significant.

#### 2.5.1. Network Parameters

The network training was scan- and subject-specific. We implemented the encoder and decoder in Pytorch 2.4.1 as fully-connected multilayer perceptrons with ReLU activations. Following previous CVAE work for motion identification[14], the encoder had four layers with 25, 50, 100, and 70 features, respectively. The decoder also had four layers with 70, 100, 50, and 25 features, respectively. The Adam optimizer was used with a learning rate of 0.001. We empirically selected the KL regularization weight $\beta$ by inspecting the shapes of the learned latent trajectories for one scan and chose the smallest $\beta$ that produced physiologic cardiac and respiratory motion while maintaining rapid convergence. We used the same $\beta$ value for all three sequences and all subjects. For frequency-based disentanglement of cardiac and respiratory motion, we set the low-pass cutoff for respiratory motion to 0.67 Hz and the band-pass range for cardiac motion to 0.67–2.5 Hz.

## 3. Results

For cine imaging, Generative Multitasking replaces discrete cardiac phase bins with a continuous variable-amplitude cardiac trajectory, revealing beat-to-beat functional variability that gated reconstructions miss (Fig. 3). Discrete cardiac phase binning produces a piecewise constant signal over the cardiac cycle and assumes that each heartbeat is functionally equivalent (aside from timing differences which can be captured during binning). In contrast, Generative Multitasking models cardiac dynamics as a continuous path in the complex harmonic latent space, yielding temporally smooth and physiologic motion across the entire cycle. This variable-amplitude representation preserves subtle beat-to-beat differences visible as heterogeneous blood pool behavior across successive heartbeats.

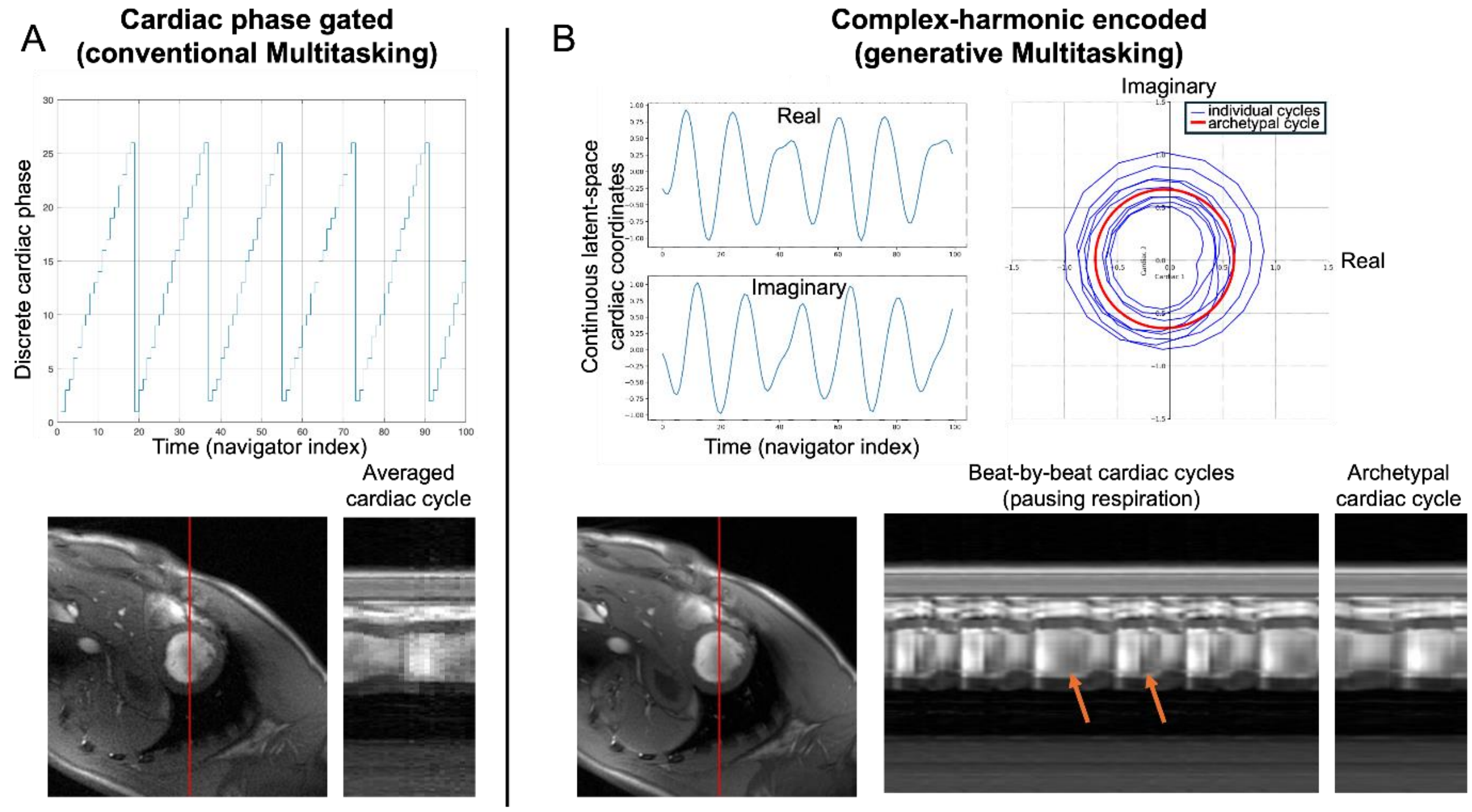

**Figure 3**: GRE cine comparison of strict cardiac phase gating (Multitasking) and complex-harmonic modeling (generative Multitasking). (A) Cardiac phase gating averages all cardiac cycles to produce a single representative cycle. (B) Complex-harmonic modeling represents the signal with real and imaginary components. Their combination into phase captures timing, as in standard gating, while the varying amplitude captures beat-to-beat differences in image behavior. After image reconstruction, the method can play back beat-to-beat cardiac behavior in the original sequence, as in real-time imaging, with the added ability to pause other effects such as respiratory motion. Red arrows indicate differences in blood pool behavior between cycles. An archetypal cycle can be defined by moving through latent space along averaged coordinates, to extract, e.g. a median cycle, without averaging together all beats.

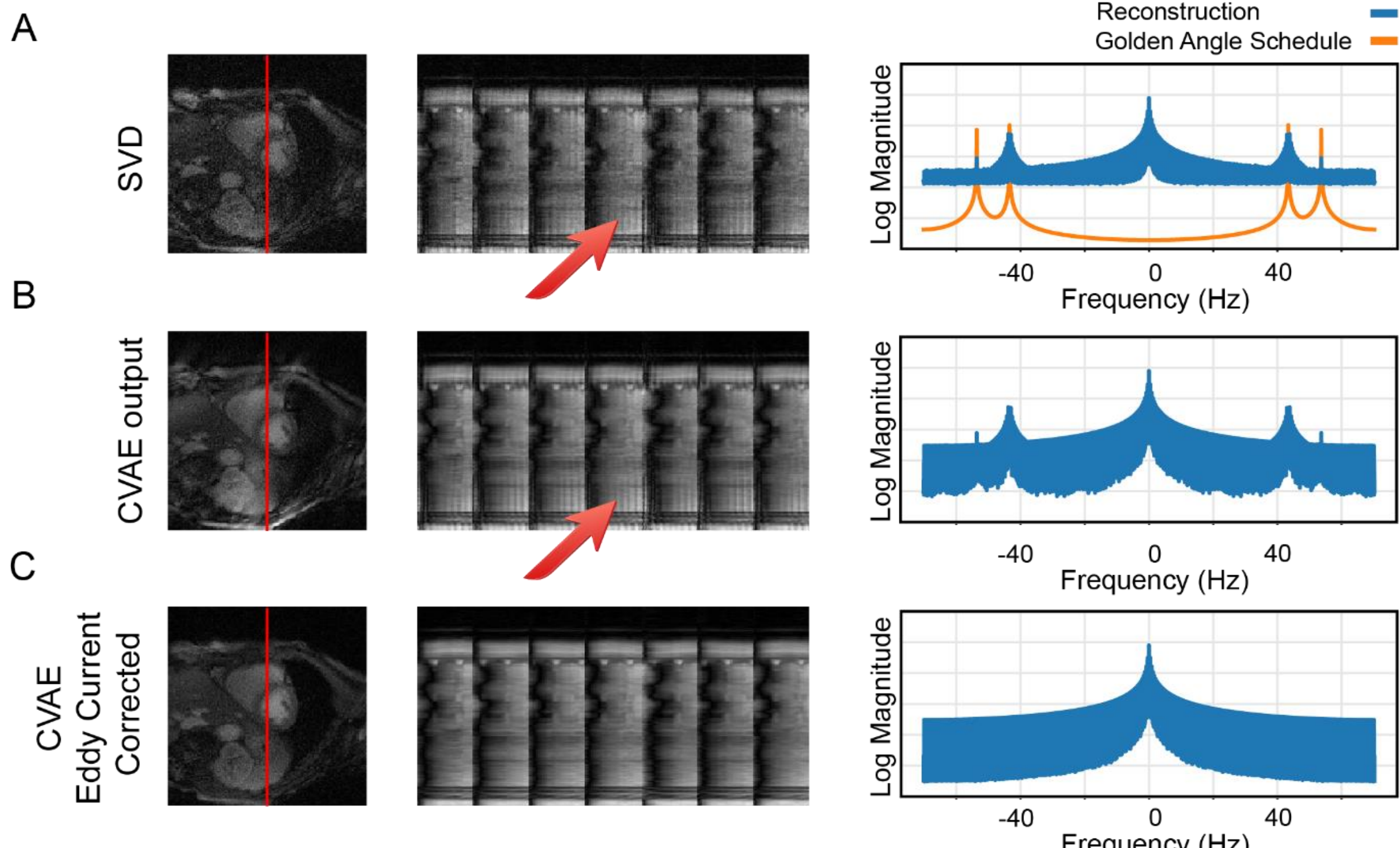

**Figure 4.** Time resolved images and their energy spectra. (A) Images from the initial SVD-based temporal subspace (input $\mathbf{\Phi}$) are noisy and show high frequency fluctuations (*red arrow*). These fluctuations match the energy spectrum of the golden-angle sampling schedule $\exp(\pm in\angle k[t])$, $n$=1,2 (*orange*). (B) A reconstruction using the CVAE output without eddy-current correction (output $\widehat{\mathbf{\Phi}}$), improves SNR but retains high-frequency fluctuations. (C) A reconstruction using the CVAE output with eddy-current correction at inference ($\mathbf{\Phi}_{\text{ec}}$) reduces both the noise and the high-frequency fluctuations, while preserving other high-frequency content not described by the sampling schedule.

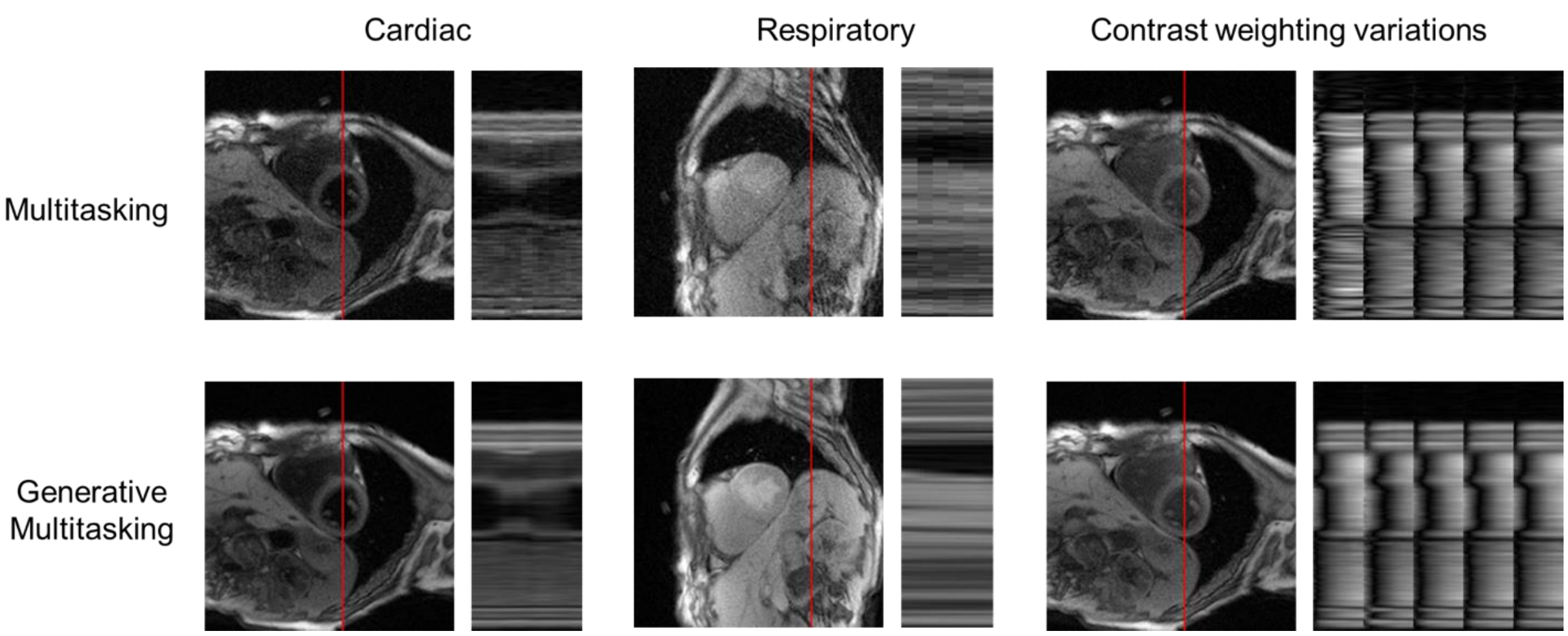

**Figure 5**. Multidimensional images from multitasking and generative multitasking. Top row is the multidimensional images from conventional multitasking. Bottom row is from generative Multitasking. The CVAE learns a continuous cardiac and respiratory motion. For display, we sampled 50 locations from cardiac coordinate and 20 from respiratory coordinate. The images from generative Multitasking preserve cardiac motion, respiratory variation, and contrast changes while demonstrating clearly higher SNR and smoother motion handling. The continuous latent representations allow interpolation of intermediate motion states, reducing temporal discontinuities from phase-binned motion-resolved imaging.

In the time-resolved domain, eddy-current correction with generative Multitasking removes golden-angle-trajectory–related artifacts (i.e., k-space jump–related artifacts; Fig. 4). The effect is visible in representative time-resolved ("real-time") images and corresponding temporal spectra reconstructed using three methods: the initial SVD-based temporal subspace (input $\mathbf{\Phi}$), CVAE without eddy-current correction at inference (output $\widehat{\mathbf{\Phi}}$), and CVAE with eddy-current correction at inference ($\mathbf{\Phi}_{\mathrm{ec}}$). The SVD reconstruction exhibits higher noise and fluctuations at high temporal frequencies that coincide with the spectrum of $\exp(\pm in\angle k[t])$, $n$=1,2, i.e., the second-order harmonic of the golden-angle-trajectory sampling schedule. The uncorrected CVAE reconstruction improves noise but retains the high-frequency fluctuations. The eddy-corrected CVAE reconstruction removes the high-frequency fluctuations while preserving other high-frequency content not described by the sampling schedule. This is further visible in Video S1. The same reconstruction can be displayed in the multidimensional domain, where generative Multitasking preserves contrast variations while improving SNR and yielding continuous cardiac and respiratory motion (Fig. 5). The continuous-valued latent space coordinates of generative Multitasking provide a more continuous representation of motion, compared with the 20 discrete cardiac phases and 6 discrete respiratory bins in the conventional Multitasking reconstruction.

In the mapping experiments, generative Multitasking produced higher-SNR T1 and T2 maps while also better preserving fine details at the blood–myocardium interface such as trabeculae and papillary muscles (Fig. 6). These improvements suggest that CVAE-based motion estimation and latent temporal representation provide a denoised temporal basis and accurate motion estimates, which in turn improve T1 and T2 maps from free-breathing, non-ECG-gated acquisitions. Across subjects and slices, the intraseptum coefficient of variation (CoV) of T1 values were $0.13 \pm 0.04$ for generative Multitasking and $0.31 \pm 0.06$ for conventional Multitasking; intraseptum T2 CoVs were $0.12 \pm 0.05$ for generative Multitasking and $0.32 \pm 0.13$ for conventional Multitasking (Fig. 7). ANOVA results confirmed that generative Multitasking produced lower intraseptum CoV (higher SNR) for both T1 ($p<0.001$) and T2 ($p<0.001$) maps.

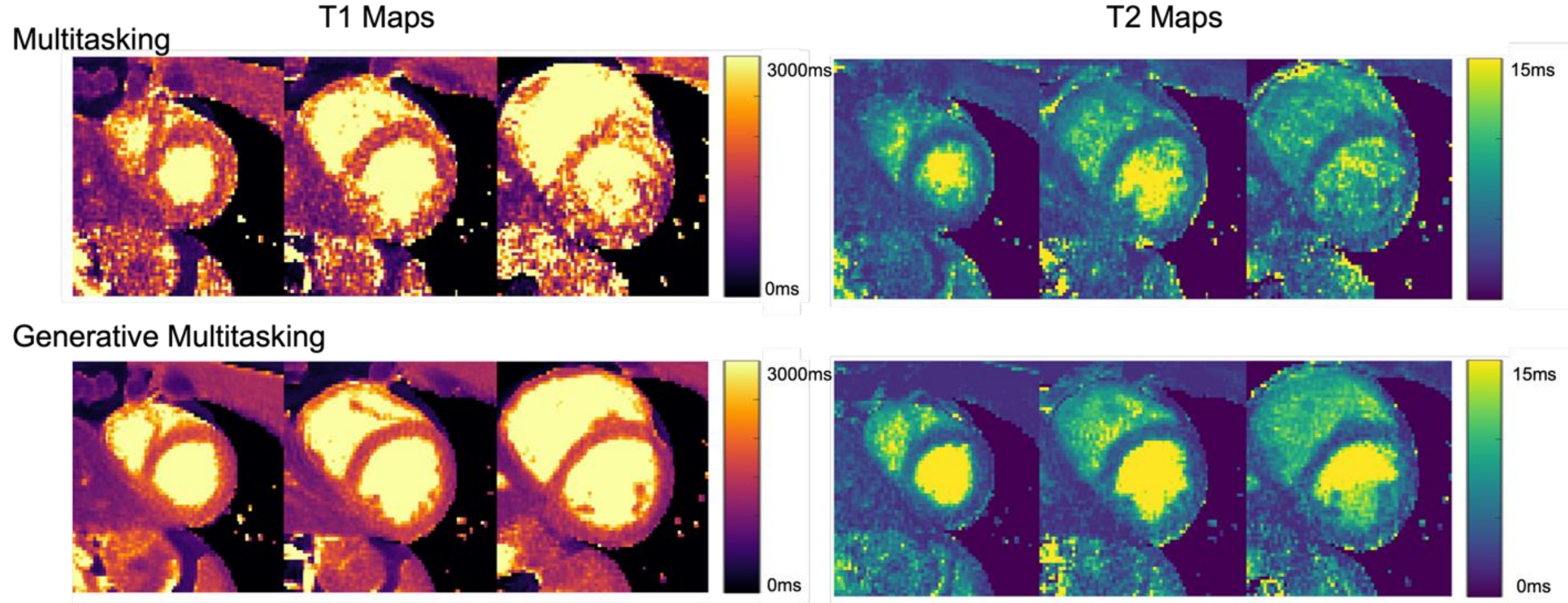

**Figure 6.** T1 and T2 maps of Multitasking and generative Multitasking from one healthy subject. Maps from generative Multitasking demonstrates visibly higher SNR comparing to conventional Multitasking while preserving details at the blood-myocardium interface.

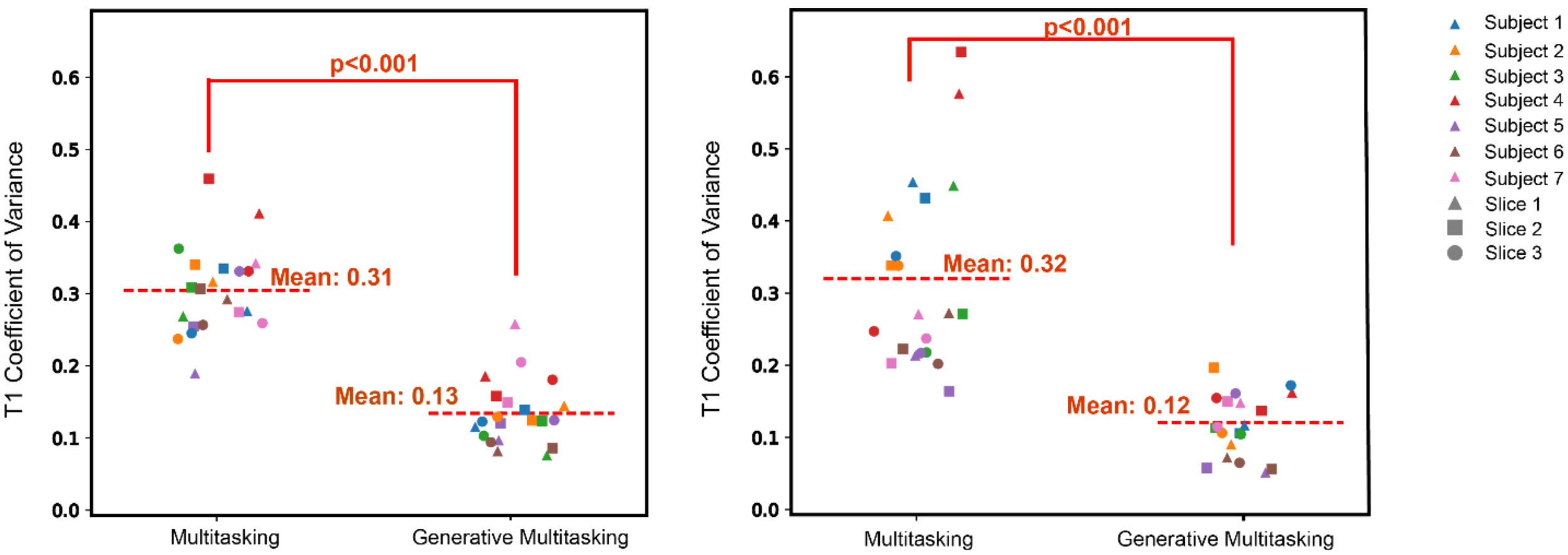

**Figure 7**. Comparison of intraseptal T1 and T2 coefficient of variation (CoV) between conventional Multitasking and generative Multitasking across subjects and slices. (Left) Intraseptal T1 CoV values for seven subjects (three slices per subject) reconstructed using conventional Multitasking (left cluster) and generative Multitasking (right cluster). Generative Multitasking demonstrates a markedly lower CoV distribution (Mean = 0.13) compared with conventional Multitasking (Mean = 0.31), indicating substantially improved signal-to-noise ratio (SNR) and temporal stability ($p < 0.001$). (Right) Intraseptal T2 CoV values, where generative Multitasking again shows a reduced CoV (Mean = 0.12) relative to conventional Multitasking (Mean = 0.32; $p < 0.001$).

For the *n*=7 scans in the quantitative experiment, conventional Multitasking's median (interquartile range [IQR]) binning time was 129 (119–149) s, and tensor subspace estimation was 709 (658–994) s, for a total of 838 (771–1169) s. For generative Multitasking, the total CVAE training and inference time was 797 (585–800) s, which was modestly but significantly shorter ($p = 0.016$).

## 4. Discussion

This work introduces a generative Multitasking framework that bridges conventional cardiac phase-gated imaging with real-time CMR by representing the cardiac cycle as a complex harmonic with both phase and a variable amplitude. Whereas cardiac phase continues to encode timing, the latent amplitude dimension provides an avenue to capture beat-to-beat functional variability, enabling a single model to span gated ("cardiac phase-resolved") and time-resolved ("real-time-like") views of cardiac motion. This softens the traditional trade-off between high-spatial-resolution gated cine and variability-robust real-time imaging: the same free-breathing, non-ECG-gated acquisition can be reconstructed as a conventional archetypal cine or as a beat-resolved, real-time-like series, with the choice deferred to reconstruction and analysis rather than prescribed at the time of scanning. The complex harmonic coordinate system further provides a continuous space in which multiple beats can be organized, interpolated, and, in principle, clustered, laying the groundwork for more structured analysis of real-time data.

In the cine experiments, the continuous latent space for cardiac and respiratory motion produced temporally smooth, physiologic trajectories while still preserving differences in behavior across heartbeats. In a healthy volunteer, the latent cardiac amplitude captured subtle beat-to-beat differences that manifested as heterogeneous blood pool dynamics, rather than enforcing the assumption of identical cycles. If this behavior generalizes to arrhythmias with some underlying temporal structure (e.g., bigeminy, premature ventricular contractions, or supraventricular tachycardia), this could provide a robust tool for imaging patients for whom CMR is often considered too challenging to perform. In these patients, the latent amplitude dimension may also offer an avenue for selecting which beat to analyze or to provide a natural axis for representing different types, an continuous-valued extension of beat-type gating or R-R interval clustering[15,16], potentially allowing separate cines of different beat types (e.g., dominant sinus beats versus ectopic beats) from a single free-breathing, non-ECG-gated scan. For more complex arrhythmias, it may be necessary to relax or remove the bandpass and harmonic constraints on the cardiac latent space and instead learn a more flexible ≥2D cardiac representation, essentially a continuous-valued, multi-contrast extension of recent clustering approaches that have been successful in steady-state cine for atrial fibrillation[17,18].

In the time-resolved multicontrast experiments, generative Multitasking behaved as an implicit correction model for trajectory-dependent system effects such as eddy-current artifacts. Conditioning the CVAE on the previous k-space angle and then replacing this term at inference with a constant removed the golden-angle-synchronized fluctuations seen in the temporal spectra of the SVD and uncorrected CVAE reconstructions. Importantly, the correction did not simply suppress high temporal frequencies as a whole; instead, it selectively reduced narrow peaks that aligned with the repetition of the golden-angle sampling pattern, while preserving other high-frequency components of motion and contrast repetitions. The SNR improvement was also seen in the multidimensional domain, where motion was preserved. This pattern suggests that the model is targeting systematic effects tied to the preceding trajectory, eddy currents, gradient system memory, and/or incomplete spoiling, not acting as a generic temporal smoother.

Quantitative mapping results further highlight the combined impact of the generative Multitasking framework on motion handling and temporal modeling. Compared with conventional Multitasking, generative Multitasking consistently produced T1 and T2 maps with lower intraseptal CoV, corresponding to roughly a twofold SNR gain while preserving fine structures at the blood–myocardium interface. This reduction in variability was achieved without the spatial regularization usually used with Multitasking, such as wavelet-sparse regularization, suggesting that the CVAE-derived temporal basis itself is better-conditioned and more accurately represents the true motion and contrast evolution.

This study should be viewed as a proof-of-concept; several limitations remain. First, validation was limited in scope: experiments were performed in healthy volunteers, and we did not conduct dedicated cine or quantitative mapping studies in larger or more diverse patient cohorts. In particular, targeted studies in patients with arrhythmias are needed to test how well the latent amplitude dimension captures truly pathological beat-to-beat variability and to establish whether a single free-breathing acquisition can reliably replace separate gated cine and real-time protocols in that setting. We based the high-level CVAE structure and parameters here on a previous study that used a CVAE for motion identification[14], which was not built for complex-harmonic cardiac space or the dual use of the CVAE as an implicit neural representation of a temporal subspace. The network architecture, disentanglement filters, and scan-specific training strategy may benefit from further optimization for this dual purpose, including alternative networks, adaptive frequency bands, time-dependent encoders/decoders, and other technical variations. Although Figure S1 illustrates that the time-dependent modules can preserve a range of timing and amplitude irregularity, we did not systematically establish the degree of cardiac or respiratory aperiodicity at which this behavior breaks down. Because the latent motion coordinates are continuous, there is no fixed finite number of respiratory states or cardiac phases in this framework, and a simple, meaningful definition of temporal resolution remains an open question that may relate to the learned variances in the latent space. We also did not compare ventricular volumetry or absolute T1/T2 values against established conventional reference techniques, so agreement with standard clinical measurements remains to be determined. Although the current study explored our new complex harmonic cardiac representation in a CVAE-based generative Multitasking setting, it is conceptually not tied to a specific network architecture or reconstruction framework. Future work should explore incorporating variable-amplitude cardiac coordinates into other contexts[19-23] to determine how broadly this representation can improve cardiac-resolved imaging.

## 5. Conclusions

In conclusion, we presented a generative Multitasking framework that bridges conventional cardiac phase-gated imaging and real-time CMR. A complex harmonic cardiac representation with both phase and latent amplitude enables a single model to capture timing as well as beat-to-beat functional variability, effectively combining cine-like depiction of anatomy and function with real-time-like robustness to irregular beats within a single free-breathing, non-ECG-gated acquisition. Using a scan-specific neural network with implicit neural representations, the generative Multitasking method learns continuous cardiac coordinates which organize multiple disparate cardiac cycles and a denoised temporal basis, which together improve the flexibility of cardiac motion representation, suppress trajectory-dependent artifacts, and improve T1 and T2 mapping.

## Code Availability Statement

Example Python CVAE code is available under an Academic Software License upon reasonable request to the corresponding author.

**Supplemental Information:**

**Video S1:** Time-resolved images reconstructed using three different temporal-subspace solutions. (A) Images from the initial SVD-based temporal subspace (input $\mathbf{\Phi}$) yields noisy images with pronounced high-frequency fluctuations. (B) A reconstruction using the CVAE output without eddy-current correction (output $\widehat{\mathbf{\Phi}}$) improves SNR, but temporal fluctuations remain visible. (C) A reconstruction using the CVAE output with eddy-current correction at inference ($\mathbf{\Phi}_{\mathbf{ec}}$) further suppresses noise and high-frequency fluctuations without temporally blurring or losing depiction of motion.

**Figure S1:** Simulated latent signals passing through the filtering and harmonic analysis modules. The bandpass filter passes through 0.67–2.5 Hz. Signals shown are: 1) a periodic signal (regular heart rate); 2) a frequency-modulated signal (variable heart rate); 3) an amplitude-modulated signal (regular heart rate with time-varying latent amplitude); 4) a frequency- and amplitude-modulated signal (variable heart rate with time-varying latent amplitude); and 5) pure Gaussian noise. Irregularity is generally preserved when the heart rate stays within the bandlimits, and the modules do not impose regularity on, e.g. random inputs.

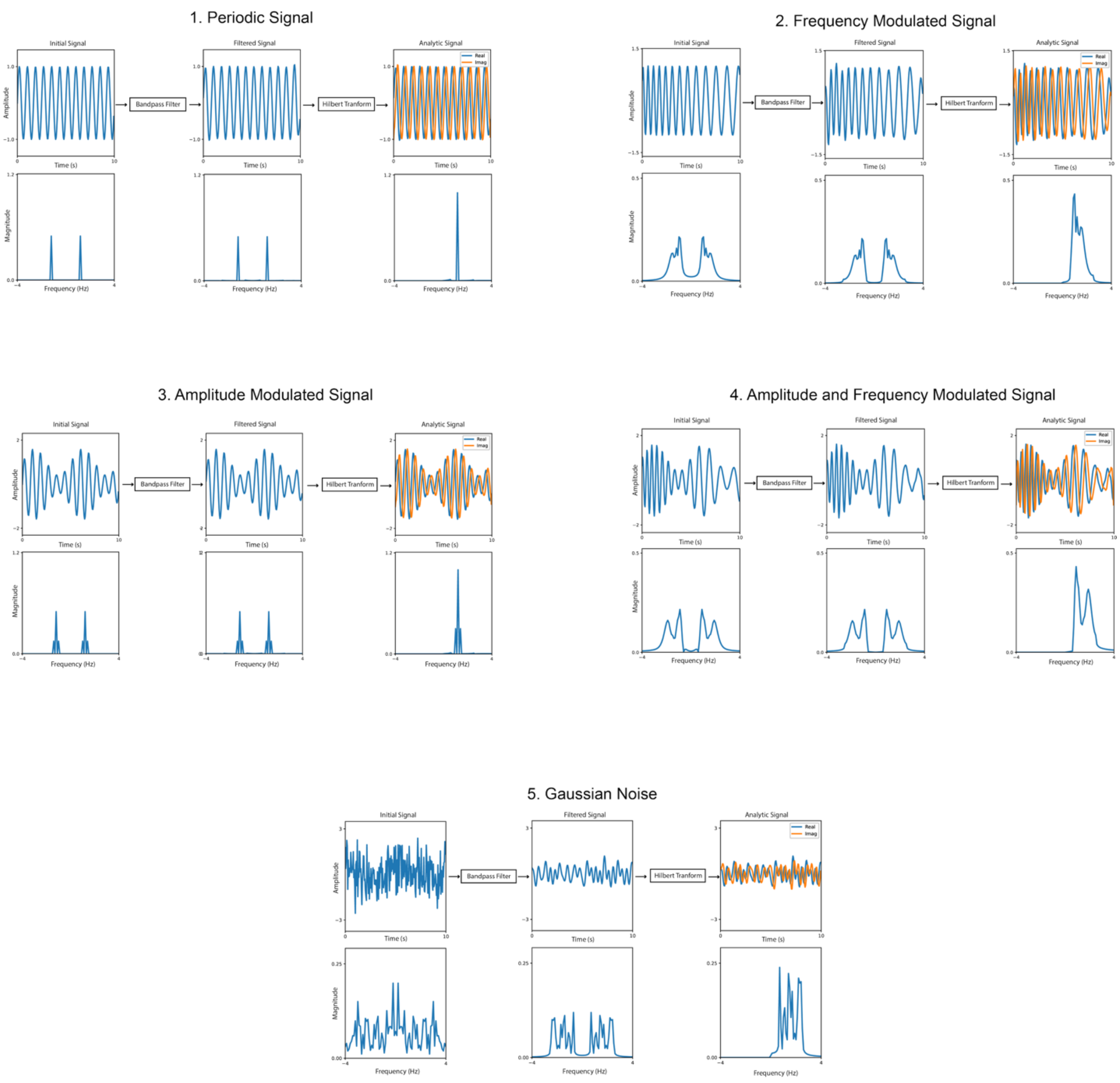

**Figure S1:** Simulated latent signals passing through the filtering and harmonic analysis modules. The bandpass filter passes through 0.67–2.5 Hz. Signals shown are: 1) a periodic signal (regular heart rate); 2) a frequency-modulated signal (variable heart rate); 3) an amplitude-modulated signal (regular heart rate with time-varying latent amplitude); 4) a frequency- and amplitude-modulated signal (variable heart rate with time-varying latent amplitude); and 5) pure Gaussian noise. Irregularity is generally preserved when the heart rate stays within the bandlimits, and the modules do not impose regularity on, e.g. random inputs.